\documentclass[12pt]{article}

\newcommand{\sgn}{\mathop{\rm sign}}
\newcommand{\rmi}{{\rm i}}
\newcommand{\be}{\begin{equation}}
\newcommand{\ee}{\end{equation}}
\gdef\journal#1, #2, #3, 1#4#5#6{{#1~}{\bf #2} (1#4#5#6) #3}
\gdef\ibid#1, #2, 1#3#4#5{{#1} (1#3#4#5) #2}

\begin{document}


\begin{center}
{\large \bf
Chern-Simons Field Theory and Generalizations of Anyons\footnote{
Contributed paper at the International Europhysics Conference on High Energy
Physics HEP-97 (Jerusalem, Israel, 19--26 August 1997).
}$^,$\footnote{Parts of this work were done in collaboration
with Edouard Gorbar, Serguei Isakov, and St\'ephane Ouvry.}
}\\[1em]
{\large
Stefan Mashkevich\footnote{mash@phys.ntnu.no}\\[1em]
\it
Bogolyubov Institute for Theoretical Physics, 252143 Kiev, Ukraine
}
\end{center}

\begin{abstract}
It is well known that charges coupled to a pure Chern-Simons gauge
field in (2+1) dimensions undergo an effective change of statistics,
i.e., become anyons.
We will consider several generalizations thereof, arising when the
gauge field is more general.
The first one is ``multispecies anyons''---charged
particles of several species
coupled to one, or possibly several, Chern-Simons fields.
The second one is finite-size anyons, which are charged
particles coupled to a gauge field described by the Chern-Simons
term {\it plus} some other term.
In fact, rigorously speaking, quasielectrons and quasiholes in the
fractional quantum Hall effect are multispecies finite-size anyons.
The third one is an analog of finite-size anyons which arises in a model with
a mixed Chern-Simons term; notably, this model is P,T-invariant,
which opens the way for practical applications even when there is no
parity-breaking magnetic field.
\end{abstract}

\section{Introduction}

The notion of statistics of identical particles has to do
with the basic principles of quantum theory.
Twenty years ago it was understood \cite{LM77}
(and later on confirmed in different ways \cite{GMS8081,W82})
that this notion
is by far more rich than one used to think. Namely, the common
statement that there can be only two types of statistics---Bose
and Fermi---is actually true only in three (or higher) dimensional
space. In two dimensions (as well as in one dimension,
which we will not discuss here), 
a continuous family of statistics is
possible, characterized by a real number---{\it statistics
parameter\/} $\alpha$, such that $\alpha=0$
corresponds to bosons and $\alpha=1$ to fermions.

Speaking a bit loosely, here is the way to see it. Imagine two
identical particles in a plane being interchanged in an anticlockwise
manner. As a result, their wave function can only pick up a phase
factor, which we will denote as $\exp[\rmi\pi\alpha]$;
without loss of generality, we can then assume $0\leq\alpha<2$.
Repeating this operation twice means, in the frame of reference
of one of the particles, that the second one is pulled all the way
around it. The corresponding phase factor is $\exp[2\rmi\pi\alpha]$.
See Fig.~1.
\begin{center}
\begin{picture}(400,90)
\thicklines
\put(80,60){\circle*{8}}
\put(80.3,63.7){\vector(-1,-3){0}}
\put(70,45){$j$}
\put(120,60){\circle*{8}}
\put(119.6,56.2){\vector(1,3){0}}
\put(122,45){$k$}
\put(100,60){\circle{40}}

\put(260,60){\circle*{8}}
\put(260.3,63.7){\vector(-1,-3){0}}
\put(250,45){$j$}
\put(280,60){\circle*{8}}
\put(280,45){$k$}
\put(280,60){\circle{40}}

\put(80,20){$\exp[\rmi\pi\alpha]$}
\put(255,20){$\exp[2\rmi\pi\alpha]$}
\put(185,5){Fig.~1.}

\end{picture}
\end{center}
Now, in three dimensions, the second operation corresponds to
``doing nothing'', because the closed line that the second particle
follows can be continuously deformed into a point.
Hence the condition $\exp[2\rmi\pi\alpha]=1$ and the usual
conclusion that $\alpha$ can only be 0 or 1. In two dimensions,
however, no such deformation is possible, so there is no
such condition and $\alpha$ can be {\it any\/} (real) number;
hence the name {\it anyons\/} \cite{W82}.

A natural question now is about the relevance of all
this to ``real life''---since all real particles {\it may}
live in three
dimensions and therefore can only be bosons or fermions.
The answer is actually positive and is provided (maybe not
uniquely) by the Chern-Simons gauge field model.
If there is a conserved current $j_{\mu}$ and a gauge field
$A_{\mu}$ and the Lagrangian of the theory is
\begin{eqnarray}
{\cal L} = {\cal L}_{G} - j_{\mu}A^{\mu} \;, \label {Eq3} \\
{\cal L}_{G} = {1 \over 2} k \epsilon^{\mu\nu\lambda}
  A_{\mu} \partial_{\nu} A_{\lambda} \;, \label{EqLG}
\end{eqnarray}
then a pointlike charge $j_{\mu} = e \delta^{0}_{\mu} \delta^{2}({\bf r})$
gives rise to a vector potential with only the angular component
different from zero,
\be
A_{\phi}({\bf r}) = - \frac{\alpha}{er}
\label{Eq5}
\ee
where  $\alpha = {e^2}/{2\pi k}, \; r = |{\bf r}|$.
In other words, a charge is at the same time a ``magnetic
flux point'', and interchanging charges gives rise to a
phase factor $\exp[2\rmi\pi\alpha]$, as in the
Aharonov-Bohm effect. Thus, {\it charge-flux composites\/}
are anyons: the coupling between charges and fluxes
``mimics'' anyon statistics.

This paper consists of three short stories,
each of which is in principle self-contained
but which are all connected by the fact that they have
to do with generalizations of the above simple picture.

\section{Multispecies anyons}

A further possibility specific for two dimensions
is to have nontrivial statistics of {\it distinguishable particles\/}
\cite{W92,dVO93,IMO95}.
Imagine that particles in Fig.~1 are not identical.
Their interchange can then change the wave function in
an arbitrary way, but pulling one around the other one
can only bring up a phase factor---because the particles come
back to their original positions. The very same considerations
as above imply that this factor has to be equal to unity in three
dimensions but can be arbitrary in two dimensions. Therefore,
{\it multispecies anyons\/} are defined as follows:
(i) an interchange of two identical particles, of species $a$,
leads to a phase factor $\exp[\rmi\pi\alpha_{aa}]$;
(ii) pulling a particle of species $b$ around a one of species $a$
leads to a phase factor $\exp[2\rmi\pi\alpha_{ab}]$.
There is a {\it statistics matrix\/} $||\alpha_{ab}||$, which
is obviously symmetric; its entries are sometimes called mutual
statistics parameters.

Charge-flux composites with different values of charges and fluxes
are multispecies anyons. In particular, so are
quasielectrons and quasiholes in the fractional quantum Hall effect.
In field theory, this is the situation with different charges
and possibly several Chern-Simons gauge fields. If there is only
one sort of charge and one Chern-Simons field, so we are still
dealing with the theory (\ref{Eq3})--(\ref{EqLG}), but different
species $a$ carry different charges $e_a$, then the statistics
matrix is given by $\alpha_{ab}=e_ae_b/2\pi k$. It is evidently
symmetric, and besides its off-diagonal elements are completely
determined by the diagonal ones, save for the sign:
$\alpha_{ab}^2 = \alpha_{aa}\alpha_{bb}$.
More generally, there may be several sorts of charges and
several gauge fields; if these are numbered by $\beta$,
the Lagrangian will be
\be
{\cal L}_{G} = \sum_\beta {1 \over 2} k^\beta
\epsilon^{\mu\nu\lambda} A_{\mu}^\beta
\partial_{\nu} A_{\lambda}^\beta -
\sum_\beta j_{\mu}^\beta A^{\beta\mu} \;.
\ee
If the $a$-th species carries charges $e_{a}^{\beta}$,
we have for the statistics matrix
\be
\alpha_{ab}=\frac{\sum_\beta e_a^\beta e_b^\beta}{2\pi k^\beta}\;.
\ee
The above connection between the diagonal and off-diagonal
elements no longer exists.

\section{Finite-size anyons}

In reality, charge-flux composites are always characterized by
a finite size. For example, in the quantum Hall effect, the
typical size of the flux is of the order of the magnetic length.
From the point of view of field theory, we are dealing with
the case \cite{M93} when the gauge field Lagrangian is not just
the Chern-Simons term but rather a sum
\be
{\cal L}_{G} =
{\cal L}_{0} + {1 \over 2} k \epsilon^{\mu\nu\lambda}
  A_{\mu} \partial_{\nu} A_{\lambda} \;.
\label{Eq6}
\ee
Most naturally, ${\cal L}_{0}$ would be the Maxwell term
$-{1 \over 4} F_{\mu\nu}F^{\mu\nu}$,
which can either be present in the theory from the
very beginning or be generated as a quantum correction \cite{15}.
There may, however, be other terms---like, for example,
$(\partial_{\mu}A^{\mu})^{2}$ in Higgs-like models \cite{16}.

Let there again be a pointlike charge in the origin. Solving the
field equations corresponding to the Lagrangian
(\ref{Eq3}), (\ref{Eq6}) yields the vector potential
$A_{\mu}({\bf r})$ produced by this charge. Assuming that
${\cal L}_{0}$ does not contain time and angle $\phi$ explicitly,
$A_{\mu}$ depends only on $r$. In the Lorentz gauge, consequently, $A_r$
vanishes; $A_0$ is responsible for a
charge-charge ``Coulomb'' interaction, which we will not
take into account (assuming, for example, the values of
the charges to be sufficiently small);
now, $A_{\phi}$, without any loss of generality, can be written
\be
A_{\phi} (r) = -\frac{\alpha}{er}\epsilon(r) \;.
\label{Eq7}
\ee
The function $\epsilon(r)$ is the form factor of the flux,
because $\Phi(r) \equiv -\frac{2\pi\alpha}{e}\epsilon(r)$
is the flux through the circle of radius $r$.
The ``statistics form factor'' is $\alpha\epsilon(r)$.
In realistic cases, $\epsilon(r)$ will be continuous and
finite everywhere; if there is no singular flux at the origin,
then $\epsilon(0)=0$, and one can normalize so that
$\epsilon(\infty)=1$.
For example, in Maxwell-Chern-Simons theory \cite{15,17}
$\epsilon(r) = 1-kr K_1(kr)$, and still $\alpha = {e^2}/{2\pi k}$.

Clearly, when the separation of anyons is much greater than
their characteristic size $R$ ($1/k$ in the example above),
they are well described by the pointlike particle model;
otherwise they are not. Assume that the bare particles are
bosons and $\epsilon(0)=0$. Then the quantum mechanical levels
tend to those of bosons when the mean separation is much less
than $R$ and to those of anyons with statistics parameter $\alpha$
when it is much greater than $R$. The equation of state, in the
high-temperature approximation, tends to that of bosons or of
anyons when $\lambda\ll R$ and $\lambda\gg R$, respectively 
($\lambda$ is the thermal wavelength) \cite{M93,M96}.
Qualitatively, this is so because in this approximation,
the partition function of the Bose or
Fermi gas coincides in the lowest order with that of the
gas of classical particles interacting via an effective potential $v(r)$
which differs considerably from zero only for
$r$ \raisebox{-1ex}{$ \stackrel{\textstyle <}{\sim} $}
$\lambda$ \cite{18}. The same holds for pointlike
ideal anyons; the effective potential for them is given by
$v(r) = -T \ln \left[ 1 + (1-2\alpha^2) \exp \left(
-2\pi r^2/\lambda^2 \right) \right]$.
Therefore, the ``effective statistics parameter''
of finite-size anyons will be $\alpha\epsilon(r)$
averaged in some way over a range of the width
of the order  $\lambda$. Evidently, that tends to 0 for
$\lambda \ll R$ and $\alpha$ for $\lambda \gg R$.

\section{Anyons in a mixed Chern-Simons model}

There is another Chern-Simons model, to some extent a hybrid
of the ones from the previous two sections, which gives rise
to anyonlike objects \cite{GM95}. This is the so-called mixed
Chern-Simons model, involving two species of fermions and
two gauge fields, with a mixed Chern-Simons term.
The Lagrangian of the model \cite{DM92} is
\be \label{e1}
{\cal L} = - \frac{1}{4g^2} f_{\mu\nu} f^{\mu\nu}
+ \bar{\psi} (\rmi \partial \hspace{-0.55em} / \hspace{0.2em}
- \tau_3 a \hspace{-0.5em} / \hspace{0.2em}
- e A \hspace{-0.5em} / \hspace{0.2em}
- \Delta) \psi - \frac{1}{4\sqrt{\partial^2}}
F_{\mu\nu} F^{\mu\nu} +
\frac{\sgn (\Delta) e}{2 \pi} \epsilon^{\mu\nu\rho}
A_{\mu} f_{\nu\rho} \;.
\ee
Here
\be \label{e2}
f_{\mu\nu} = \partial_{\mu} a_{\nu} - \partial_{\nu} a_{\mu} \; ,
\; F_{\mu\nu} = \partial_{\mu} A_{\nu} - \partial_{\nu} A_{\mu} \; ,
\ee
$A_{\mu}$ is the electromagnetic field and $a_{\mu}$  the
so-called statistical gauge field.
The two species of fermions are unified in a four-component bispinor
$\psi = \left( \begin{array}{c} \!\! \psi_1 \!\! \\
\!\! \psi_2 \!\! \end{array} \right)$,
and the interesting feature of the model is that its mass term
$\Delta \bar{\psi} \psi = \Delta \bar{\psi_1} \psi_1 -
\Delta \bar{\psi_2} \psi_2 $ is P and T-invariant (see \cite{DM92}
for details).

The field equations corresponding to the
Lagrangian (\ref{e1}) read
\be \label{e6}
\begin{array}{rcl}
 \frac{1}{g^2} \partial_{\mu} f^{\mu\nu} +
 \frac{se}{2\pi} \epsilon^{\nu\mu\lambda} F_{\mu\lambda} &
 = & j_3^{\nu} \;, \\  \\
 \frac{1}{\sqrt{\partial^2}} \partial_{\mu} F^{\mu\nu} +
 \frac{se}{2\pi} \epsilon^{\nu\mu\lambda} f_{\mu\lambda} &
 = & j^{\nu} \;,  \end{array}
\ee
where
\be \label{e7}
j^{\nu} = \bar{\psi} \gamma^{\nu} \psi \;,\qquad
j_3^{\nu} = \bar{\psi} \gamma^{\nu} \tau_3 \psi \;,\qquad
s = \sgn(\Delta) \;;
\ee
Their solutions in the Lorentz gauge for pointlike sources
(a purely quantum mechanical treatment is enough for our
purposes)
\be \label{e14}
j^{\mu}(x) = ne \delta_0^{\mu} \delta^2 (\vec{x}) \; , \qquad
j_3^{\mu} (x) = n_3 \delta_0^{\mu} \delta^2 (\vec{x})
\ee
are \cite{GM95}
\be \label{e16}
A_{\phi} (r) = - \frac{sn_3}{2er} v(fr)  \;, \qquad
a_{\phi} (r) = - \frac{sn}{2r} v(fr) \;,
\ee
where $f=(eg/\pi)^2$ and
\be \label{e17}
u(x) = {\bf H}_0 (x) - Y_0 (x) = \frac{2}{\pi} x \, {}_1 F_2
\left( 1; \frac{3}{2} , \frac{3}{2} ; -\frac{x^2}{4} \right)
- Y_0 (x) ,
\ee
\be \label{e18}
v(x) = x \int_0^{\infty} \exp (-x \sinh t - t) \, dt \, .
\ee
We again neglect the ``Coulomb'' interaction.

Now imagine a composite of $n_1$
fermions of the first sort ($\psi_1$) and $n_2$ of the
second sort ($\psi_2$). According to (\ref{e7}) and (\ref{e14}),
$n = n_1 + n_2$ and $n_3 = n_1 - n_2$.
The phase factor arising from an interchange of two bare
composites is $\exp[\rmi\pi n^2]$, consequently
they are fermions/bosons for odd/even $n$.
However, the potentials (\ref{e16}) lead to a change of statistics.
The ``statistics form factor'' as defined above is
\begin{eqnarray}
\alpha\epsilon(r) & = & - \frac{1}{2\pi}
 \left[ ne A_{\varphi} (r) \cdot 2 \pi r +
 n_3 a_{\varphi} (r) \cdot 2 \pi r \right] \nonumber \\
 & = & snn_3 v(fr) = s (n_1^2 - n_2^2) v(fr) \;. \label{e19}
\end{eqnarray}
The function $v(x)$ equals 0 at $x=0$ and tends to 1
at $x\to\infty$. Therefore, at distances much greater
than $f^{-1}$ the phase factor is
$\exp[\rmi\pi((n_1 + n_2)^2 +s(n_1^2 - n_2^2))]=1$
for any integer $n_1$  and $n_2$. Thus,
at large distances the composites in question always
behave as bosons. At intermediate distances,
they behave like finite-size anyons.
This model is interesting, as mentioned before, because
of its P and T-invariance; therefore its relevance to
real systems appears considerably more probable than
for the usual P and T-noninvariant Chern-Simons model.

\section{Conclusion}

We have considered three different but to some extent
interconnected generalizations of anyons, which arise
from coupling charges to Chern-Simons-like gauge fields
of different forms. Ideal (i.e., single-species pointlike)
anyons arise as particular or limiting cases.
One expects that the generalizations considered will describe
the experimental situation more exactly than the ideal
anyon model; it is therefore interesting to study them
in more detail.


\begin{thebibliography}{99}

\bibitem{LM77} J.M. Leinaas and J. Myrheim,
  \journal Nuovo Cim. B, 37, 1, 1977.
\bibitem{GMS8081} G.A. Goldin, R. Menikoff, D.H. Sharp,
  \journal J. Math. Phys., 21, 650, 1980;
  \journal {\it ibid.}, 22, 1664, 1981.
\bibitem{W82} F. Wilczek,
  \journal Phys. Rev. Lett., 49, 957, 1982.
\bibitem{W92} F. Wilczek,
  \journal Phys. Rev. Lett., 69, 132, 1992.
\bibitem{dVO93} A. Dasni\`eres de Veigy and S. Ouvry,
  \journal Phys. Lett. B, 307, 91, 1992.
\bibitem{IMO95} S.B. Isakov, S. Mashkevich, and S. Ouvry,
  \journal Nucl. Phys. B, 448, 457, 1995.
\bibitem{M93} S. Mashkevich,
  \journal Phys. Rev. D, 48, 5953, 1993.
\bibitem{15} K. Shizuya, H. Tamura,
  \journal Phys. Lett. B, 252, 412, 1990.
\bibitem{16} X.C. Wen, A. Zee,  preprint NSF-ITP-88-114 (1988).
\bibitem{17} S. Mashkevich, H. Sato, G.M. Zinovjev,
  \journal JETP, 78, 105, 1994.
\bibitem{M96} S. Mashkevich,
  \journal Phys. Rev. D, 54, 6537, 1996.
\bibitem{18} K.Huang, {\em Statistical Mechanics}, John Wiley
  \& Sons, New York- London (1963).
\bibitem{GM95} E. Gorbar, S. Mashkevich,
  \journal Z. f. Phys., C65, 705, 1995.
\bibitem{DM92} N. Dorey, N.E. Mavromatos,
  \journal Nucl. Phys. B, 386, 614, 1992.

\end{thebibliography}
\end{document}